\documentclass[12pt,preprint]{aastex}
\usepackage[numberedappendix]{emulateapj5}
\usepackage{amstext}
\usepackage{apjfonts}
\usepackage{amsmath}

\journalinfo {\underline{\bf To Appear in Astrophysical Journal Letters.}}
\shorttitle{SHOCKED BLACK HOLE ACCRETION FLOW AND QPO}
\shortauthors{DAS}

\begin{document}


\title{Role of shocked accretion flows in regulating the QPO
of galactic black hole candidates}


\author{Tapas K. \ Das$^{1,2}$}
\affil{$^1$Division of Astronomy, Department of Physics and Astronomy,
University of California at Los
Angeles, Box 951547, Los Angeles, CA 90095-1562, \\
$^2$ Institute of Geophysics and Planetary Physics, 
University of California at Los Angeles, 
Box 951567, Los Angeles, CA 90095-1567, USA\\
e-mail: tapas@astro.ucla.edu}
\begin{abstract}
\noindent
Using a generalized non-spherical, multi-transonic
accretion flow model, we analytically
calculate the normalized QPO frequency ${\overline {\bf {\nu}}}_{qpo}$ of
galactic black hole candidates in terms of 
dynamical flow variables 
and self-consistently study the dependence of
${\overline {\bf {\nu}}}_{qpo}$ on such variables.
Our results are in fairly close agreement with 
the observed QPO frequencies of GRS 1915+105.
We find that ${\overline {\bf {\nu}}}_{qpo}$  is quite sensitive
to various parameters describing the black hole accretion
flow containing dissipative and 
non-dissipative shock waves. Thus
the QPO phenomena is, {\it indeed}, regulated by `shocked' black hole accretion,
and, for the first time, we establish a definitive connection between the
QPO frequency and the properties of advective BH accretion flows.
This information may provide 
the explanation of some important observations 
of galactic micro quasars.
\end{abstract}


\keywords{accretion, accretion disks --- black hole physics ---
hydrodynamics --- shock waves --- QPO}


\section{Introduction}
\noindent
It has been established in recent years that a number of galactic black hole (BH)
candidates
show quasi-periodic oscillation (QPO) behaviours spanning a wide range of frequency
starting from a few tens of milliHertz to a few hundred Hz
(see Cui 1999 for a review). The characteristic features of these 
QPOs suggest that such
oscillations are a diagnostic of the accretion processes in the inner regions of the
accretion discs around the central compact objects, like shock
oscillations in sub-Keplerian advective accretion discs
(Chakrabarti \& Titarchuk 1995, CT hereafter,
Molteni, Sponholz \& Chakrabarti 1996, MSC hereafter,
Kazanas, Hua \& Titarchuk 1997, Titarchuk, Lapidus \& Muslimov 1998,
Chakrabarti \&
Manickam 2000, CM hereafter).
The galactic
microquasar GRS 1915+105 shows
several properties which indicate that
QPOs may be generated because of the oscillation of the hot, dense
inner regions of 
black hole accretion discs (Paul et al. 1998, Muno, Morgan \& Remillard 1999,
Yadav et al. 1999, Rutledge et al.
1999,
Smith, Heindl \& Swank 2002).
Numerical simulation of MSC revealed the fact that for shocked 
non-spherical advective BH accretion flow, if the inflow and
cooling parameters
are properly tuned, the post-shock cooling
time becomes comparable to the advection time-scale from the shock.
A shocked flow may continuously
try to adjust
itself, resulting in  the quasi-periodic radial oscillation 
of the shock around its mean steady-state solution, with an oscillation
frequency roughly
equal to the reciprocal of the post-shock advection time scale $\tau_{adv}$.
Such oscillations do not depend on the details of the cooling mechanism
(CM) and may cause the quasi-periodic variation of the luminosity emerging 
out of the post-shock region of the accretion disc by dynamically modulating 
the size and the radiative properties of the Comptonizing cloud (CT).
One thus expects that the shock formation in 
multi-transonic advective BH accretion discs may play a crucial role
in producing QPOs.\\
In this letter,
we
study how sensitive shock oscillation is to
the parameters governing the BH accretion,
which has never been done in any of the works existing in the 
literature.
Our main motivation in this letter is to
provide a generalized
self-consistent computation of various shock
parameters ${\cal P}_{sh}$\footnote {In this work, any quantity changing discontinuously at
shock location is termed as shock parameters.} in terms of fundamental inflow 
variables ${\cal P}_{ace}$ for accretion in {\it all} available
post-Newtonian pseudo-Schwarzschild BH potentials, and then to use these parameters 
to estimate the value of $\nu_{qpo}$. We also study the dependence of $\nu_{qpo}$ on
all ${\cal P}_{acc}$ and ${\cal P}_{sh}$.
We assume that the shock solutions we obtain in this letter, satisfy
the criteria of fine tuning inbetween the post-shock advection and cooling
time scales  so that
shocked flow oscillates with a
frequency $\sim1/{\tau_{\rm adv}}$ even for realistic flows with all dissipative
terms included.
Our generalized formalism assures
that our results are not just an artifact of a particular type of potential only
and inclusion of every BH potential allows
a substantially extended zone of parameter space
allowing for the
possibility of shock induced QPOs. Also, there are possibilities that in future someone
may come up with a pseudo-Schwarzschild potential better than any of the BH potentials
available in the literature, which will be the best approximation for complete general relativistic
investigation of multi-transonic shocked flow. In such case, if one
already formulates a generalized model for multi-transonic shocked
accretion disc
for any arbitrary potential, exactly what we have done in this paper,
then that generalized model will be able to readily accommodate that new potential
without requiring any significant change in the fundamental
structure of the formulation and solution scheme of the model
and we need not
worry about providing any new scheme exclusively valid only for
that new potential, if any.
\section{Governing equations, solution procedure and results}
Following are the four pseudo-potentials on which we concentrate in this work:
$$
\Phi_{1}(r)=-\frac{1}{2(r-1)}~;~
\Phi_{2}(r)=-\frac{1}{2r}\left[1-\frac{3}{2r}+12{\left(\frac{1}{2r}\right)}
^2\right]
$$
$$
\Phi_{3}(r)=-1+{\left(1-\frac{1}{r}\right)}^{\frac{1}{2}}
~;~\Phi_{4}(r)=\frac{1}{2}ln{\left(1-\frac{1}{r}\right)}
\eqno{(1)}
$$
where $r$ is the radial co-ordinate scaled in units of Schwarzschild radius 
$r_g$ (=$2GM_{BH}/c^2$). Throughout this paper
the radial distances and velocities are scaled in units of $r_g$ and $c$
respectively and all other derived quantities are scaled accordingly. Also
$G=c=M_{BH}=1$ is used.
In eq. (1),
$\Phi_1$ and $\Phi_2$ had been proposed by Paczy\'nski
\&  Witta (1980) and Nowak \& Wagoner
(1991) respectively, and
$\Phi_3$ and $\Phi_4$ are suggested by Artemova et al (1996).
For these potentials,
Das (2002, D02 hereafter) and Das, Pendharkar \& Mitra (2003, DPM
hereafter) have provided a unified approach for shock solutions in multi-transonic
black hole accretion disc for non-dissipative shocks in polytropic flow (D02) as well as
in dissipative shocks in isothermal flows (DPM). 
We follow the formalism presented in D02 and DPM 
to calculate ${\cal P}_{sh}$ for evaluating $\nu_{qpo}$.
\subsection {Polytropic accretion ; Rankine-Hugoniot shock}
\noindent
For this case, 
${\cal P}_{ace}{\equiv}
\left\{{\cal E},\lambda,\gamma\right\}$, where ${\cal E}$ is
the specific energy of the flow, $\lambda$ is the
specific flow angular momentum, and $\gamma$ is the adiabatic index of the flow.
We calculate
the shock location $r_{sh}$, along with other shock parameters like shock
strength ${\cal S}_i$(=$M_{-}/M_{+}$, $M$ being the Mach number of the flow), shock
compression ratio $R_{comp}$(=${\rho}_{+}/{\rho}_{-}$, $\rho$ being the
local matter density of accretion disc, related to its vertically integrated 
value $\Sigma$ as $\rho=\Sigma/I_nh(r)$, $h(r)$ being the local disc height and
$I_n=\left(2^nn!\right)^2/\left(2n+1\right)!$, $n=1/(\gamma-1)$, see 
Matsumoto et al. 1984) and entropy
enhancement rate at shock $\beta$ (=${\dot {\cal M}}_{+}/{\dot{\cal M}}_{-}$,
${\dot {\cal M}}$ being the entropy accretion rate) as a function of
${\cal P}_{ace}$.
For low $\lambda$ flow with moderate or strong shocks, the post-shock 
velocity is proportional to $R_{comp}r_{sh}^{3/2}$ (see also CM) and one obtains:
$$
\nu_{qpo}{\sim}\frac{1}{\tau_{adv}\left(\sim{r_{sh}}/u_{sh}\right)}=
\frac{\cal A}{R_{comp}r_{sh}^{3/2}}
\eqno{(2)}
$$
where ${\cal A}$ is a scaling constant. 
Let $\Pi$ be any physical quantity which changes discontinuously at the shock. We
define the ratio of the post to pre-shock value of $\Pi$ to be 
$
{\Re}_{\Pi}=\frac{{\Pi}_{+}}{{\Pi}_{-}}
$. We multiply the numerator and denominator of eq. (2) by $R_{comp}$ and use the 
following relations:
$$
p=\frac{\cal W}{I_{n+1}h(r)},~R_{comp}=
{\cal S}_i{{\Re}_\theta}^{-1}
=\left(\frac{p_+}{p_-}\right){{\Re}_\theta}^{-1},~
\frac{\gamma-1}{2}M^2=\frac{{\cal E}_{\rm kinetic}}{{\cal E}_{\rm thermal}}={\cal E}^{k}_{th}
\eqno{(3)}
$$
to obtain the QPO frequency as:
$$
{\bf {\nu}}_{qpo}={\cal A}
\frac{{\Re}_{\dot {\cal M}}{\Re}_{\Sigma}^2{\Re}_{\theta}^{\frac{3\gamma-5}{2\left(\gamma-1\right)}}}
{{\Re}^{\frac{1}{2}}_{{\cal E}^{K}_{th}}{\Re}_{\cal W}r_{sh}^{\frac{3}{2}}}
\eqno{(4)}
$$
 where ${\cal W}$ and $\theta$ are the vertically integrated fluid pressure and
 the flow temperature respectively. ${\cal E}^{K}_{th}$ is the ratio of flow
 kinetic energy to the total thermal energy content of the flow. 
All relevant ${\Pi}$ and
 ${\Re}_{\Pi}$ could be interpreted as ${\cal P}_{sh}$. For
 a particular set of ${\cal P}_{acc}$
for which shock forms for any specific $\Phi_i$, {\it all} ${\Re}_{\Pi}$s
 could be computed in terms of
 ${\cal P}_{ace}$ for {\it all}
available $\Phi_i$s, hence we claim that the
 expression for ${\nu}_{qpo}$ in eq. (4) is more general compared
 to the QPO frequency obtained by any other previous work. A suitable choice of
${\cal A}$ provides the maximum value of ${\bf {\nu}}_{qpo}$ (for any ${\cal P}_{acc}$
in all $\Phi_i$) to be 67 Hz for one Eddington accretion rate onto a 10$M_{\odot}$ BH,
which is the representative case for GRS 1915+105.
Hereafter, we define the normalized QPO frequency as ${\overline{\bf {\nu}}}_{qpo}=
{\bf {\nu}}_{qpo}/67$.\\
In figure 1., we plot
 ${\overline{\bf {\nu}}}_{qpo}$ as a function of various fundamental 
 ${\cal P}_{ace}$. While the
${\overline{\bf {\nu}}}_{qpo}$ is plotted along Z axis
 (denoted by $\nu$ in the figure), specific energy $\cal E$ and specific  angular
 momentum $\lambda$ are plotted along X and Y axis respectively for
$\Phi_1$(A), $\Phi_2$(B), $\Phi_3$(C) and $\Phi_4$(D).
Each surface represents the plot for a fixed $\gamma$. The left most surface is 
drawn for $\gamma=\frac{4}{3}$ and successive surfaces to the right 
are drawn for
$\left(\gamma+n{\Delta}{\gamma}\right)$ where ${\Delta}{\gamma}=0.1$
and $n$ is an integer. For $\Phi_1$ as well as for $\Phi_4$, we obtain the figure
for $n$=1, 2; while for $\Phi_3$ we use $n$=1 and for $\Phi_2$ we use $n$=0.
It is observed from the figure that 
${\overline {\bf {\nu}}}_{qpo}$ anti-correlates with
$\lambda$ and $\gamma$. This indicates that the
ultra-relativistic flow 
\footnote{By the term `ultra-relativistic' and `purely non-relativistic'
we mean a flow with
$\gamma=\frac{4}{3}$ and $\gamma=\frac{5}{3}$ respectively,
according to the terminology used in
Frank et. al. 1992.}
produces stronger
QPO compared to the purely-non relativistic flows when
other initial parameters are unchanged. 
Weakly rotating flows are better
candidates for producing high frequency oscillation;
this is due to the fact that the shock location
anti-correlates with  
${\overline {\bf {\nu}}}_{qpo}$ but correlates with
$\lambda$ (see Fig. 2). The lower the be the value of angular
momentum, the closer will be the post shock region to
the event 
horizon and more amount of the gravitational energy
will be available to dump on to the oscillation energy
in the post shock region. 
For most of the cases,
${\overline {\bf {\nu}}}_{qpo}$ correlates with $\cal
E$ while ${\overline {\bf {\nu}}}_{qpo}$
anti-correlates with $\cal E$ for relatively fewer
cases.
Such a ${\overline {\bf {\nu}}}_{qpo}-{\cal E}$
(anti) correlation are supported by observational
evidences that most QPOs strengthen towards higher
energies
(van der Klis 1995; Cui et al. 1997;
Morgan et al. 1997; Remillard et al. 1998), although some follow the
opposite trend (Morgan et al. 1997; see also Remillard et al. 1999 and
Rutledge et al. 1999 for other possible cases). 
If
${\overline{\nu}}_{qpo}{\bigg{\vert}}_{\Phi_i}^{max}$ represents the maximum
possible value of normalized QPO frequency for a particular astrophysical
source for any specific $\Phi_i(r)$, we find that:
$$
{\overline{\nu}}_{qpo}{\Bigg{\vert}}_{\Phi_3}^{max}~>~
{\overline{\nu}}_{qpo}{\Bigg{\vert}}_{\Phi_4}^{max}~>~
{\overline{\nu}}_{qpo}{\Bigg{\vert}}_{\Phi_1}^{max}~>~
{\overline{\nu}}_{qpo}{\Bigg{\vert}}_{\Phi_2}^{max}
\eqno{(5)}
$$
For ultra-relativistic shocked flow in certain range of 
${\cal E}$ and $\lambda$, 
in Fig. 2, we plot ${\overline{\bf {\nu}}}_{qpo}$
(plotted along the Y axis and denoted by $f_{norm}$ in the figure) as a function
of various ${\cal P}{sh}$ for all four $\Phi_i$ marked in the figure. To plot 
in the same figure, the post shock temperature $T_{+}$ (dotted line) and the
post shock vertically integrated density $\Sigma_{+}$ (long dashed line) is
scaled as $T_{+}~\rightarrow~a_i T_{+}$ and ${\Sigma}_{+}~\rightarrow~b_i
{\Sigma}_{+}$ where $a_i$ and $b_i$ are scaling constants different for 
four different $\Phi_i(r)$.
Similar figures could also be drawn for 
{\it any} range of $\left\{{\cal E},\lambda\right\}$ for 
which a shocked flow physically exists.
It is evident from the figure that 
${\overline{\bf {\nu}}}_{qpo}$ non-linearly
anti-correlates with the shock location whereas it non linearly correlates with
the post shock temperature and post shock density. This result may have
important consequences in connection to the
jet formation mechanism for galactic relativistic sources. 
It has been shown that  (Das 1998, Das \& Chakrabarti
1999) the hot and dense post shock region of black hole accretion disc may
generate accretion-powered galactic and extra-galactic outflows and the barionic
load of such outflows increases with the increment of post shock flow
temperature and density. From Fig. 2, we conclude that there
must be some relationship inbetween the normalized QPO frequencies and the amount
of baryonic content of outflows from galactic microquasars; details of such
investigation is reported elsewhere (Das, Rao \&  Vadewale 2003).\\
In very high states of the QPOs in LMXB, radiation drag might play a
crucial role in accretion close to the event horizon and the QPO might
be caused by the radiatively driven oscillations in a quasi-spherical
accretion flow (Fortner et al. 1989). The ${\overline{\bf {\nu}}}_{qpo}-T_+$
co-relation presented in Fig. 2 is thus expected to support such QPO states
because the quasi-spherical
post-shock region in our model possesses more thermal energy as it
forms closer to the event horizon.
\subsection{Isothermal accretion; dissipative shocks}
\noindent
For such flow, 
we calculate all ${\cal P}_{sh}$s and ${\Re}_{\pi}$ in terms of 
${\cal P}_{acc}{\equiv}\left\{\lambda,T\right\}$, where $T$ is
the constant flow temperature.
The amount of energy dissipation at shock may become as much as five to
fifteen (at most) percent of the rest mass energy of the accreting material and 
may be computed as:
$$
{\Delta}\epsilon = \frac{1}{2}\left(M_{-}^{2} - M_{+}^{2} \right) -
\frac{\kappa{T}}{{\mu}m_H}
ln{{\Re}_{\Sigma}}
\eqno{(6)}
$$
This huge amount of dissipated energy may be used to power the QPO and/or 
to produce and power the 
strong X-ray flares from the BH environment.
We can calculate the normalized QPO frequency for this case as:
$$
{\overline{\bf {\nu}}}_{qpo}=
\frac{
{\rm exp}\left[\frac{{\Delta}{\epsilon}\mu{m_H}}{\kappa{T}}
-\frac{{\dot M}_{in}^2\Phi_i^{\prime}{\big{\vert}}_{r=r_{sh}}} 
{2p_+^2r^3_{sh}}
\left(1-{\Re}_{\Sigma}\right)\right]
}
{{\Re}^2_{\Sigma}r_{sh}^{\frac{3}{2}}}
\eqno{(7)}
$$
In figure 3. we plot the variation of ${\overline{\bf {\nu}}}_{qpo}$
(plotted along the Z axis and is marked as $\nu$ in the figure) with specific
angular momentum $\lambda$ (plotted along X axis) and flow temperature (plotted
along Y in the unit of $T_{10}$ (=T/10$^{10}$ Degree $K$)
and is marked as $\tau$ in the figure) for four
pseudo potentials 
$\Phi_1(A)$, $\Phi_2(B)$, $\Phi_3(C)$ and $\Phi_4(D)$. 
It is observed from the
figure that ${\overline{\bf {\nu}}}_{qpo}$ anti-correlates with $\lambda$ and
correlates with $\tau$, and:
$$
{\overline{\nu}}_{qpo}{\Bigg{\vert}}_{\Phi_3}^{max}~>~
{\overline{\nu}}_{qpo}{\Bigg{\vert}}_{\Phi_4}^{max}~>~
{\overline{\nu}}_{qpo}{\Bigg{\vert}}_{\Phi_1}^{max}~>~
{\overline{\nu}}_{qpo}{\Bigg{\vert}}_{\Phi_2}^{max}
\eqno{(8)}
$$
which is exactly same as that of polytropic flow, see eq. (5).
This clearly indicates that 
shocked accretion flow in $\Phi_3$ generates the most intensely
oscillating QPOs. This indicates that the high frequency QPOs 
observed  for any astrophysical sources may, perhaps, be better explained using
$\Phi_3$, while $\left(\Phi_1,\Phi_4\right)$ and $\Phi_2$ are 
appropriate to explain the intermediate and low frequency QPOs, respectively.
For shocked flow in certain range of $\lambda$ and $T$,
figure 4 represents the variation of ${\overline{\bf {\nu}}}_{qpo}$ (plotted
along the Y axis and is denoted by $f_{norm}$ in the figure) with 
shock location $r_{sh}$ (solid line), shock compression ratio
${\Re}_{\Sigma}$ (dotted line) and energy dissipation at shock ${\Delta}{\epsilon}$
(long dashed line).
Similar figures could also be drawn for
{\it any} range of $\left\{\lambda,T\right\}$ for
which a shocked flow physically exists.
In the figure, ${\Re}_{\Sigma}$  and ${\Delta}{\epsilon}$ are scaled as
${\Re}_{\Sigma}~\rightarrow~c_i {\Re}_{\Sigma}$ and ${\Delta}{\epsilon}~\rightarrow~d_i
{\Delta}{\epsilon}$ where $c_i$ and $d_i$ are scaling constants different for
four different $\Phi_i(r)$.
It is observed that ${\overline{\bf {\nu}}}_{qpo}$ non-linearly anti-correlates
with the shock location which indicates the generation of 
stronger oscillations closer to the event
horizon as is observed in reality (Cui 1999). Also ${\overline{\bf {\nu}}}_{qpo}$
correlates with ${\Re}_{\Sigma}$ and ${\Delta}{\epsilon}$ which indicates that the
stronger is the shock (as ${\Re}_{\Sigma}={\cal S}_i$, ${\cal S}_i$ being the shock
strength, see DPM), more intense is the oscillation and higher is the amount
of dissipated energy at the shock locations. Co-relation in between 
${\overline{\bf {\nu}}}_{qpo}$ and ${\Delta}{\epsilon}$ for isothermal 
flow may help in
understanding the characteristic features of the duration of the quiescent state
of the transient X ray source GRS 1915+105 (Das \& Rao, in preparation).
\section{Conclusion}
In this work we have examined the dependence of normalized
QPO frequencies
${\overline{\bf {\nu}}}_{qpo}$ on various parameters describing the 
generalized
multi-transonic BH accretion flow with dissipative and non-dissipative 
shock waves. 
We found that ${\overline{\bf {\nu}}}_{qpo}$ is quite sensitive
on most of the parameters responsible for producing the
post shock flow, the anticipated oscillation of which may resulting the
QPOs. This {\it indicates} that the QPOs in 
galactic sources are {\it inherently controlled} by shock dominated 
accretion flows. The fact that ${\overline{\bf {\nu}}}_{qpo}$ is sensitive to various
flow parameters found in this work reinforces the belief
that our calculation may
be useful in future to understand various observational
features of QPO in galactic BH candidates, a few of which
have already been addressed in \S 2.1 and \S 2.2. 
We performed our calculations in a generalized way by incorporating 
all available pseudo-Schwarzschild black hole potentials. However, one limitation of
the work presented here is that we could not postulate any
observational method that might convincingly discriminate among our results obtained using 
the various potentials.
\acknowledgments
\noindent
My research at UCLA
is supported by Grant No. NSF AST-0098670.
I acknowledge 
useful discussions with M. R. Morris, M. P. Muno and
A. R. Rao. 
I thank the anonymous referee for providing useful suggestions to improve the 
presentation of this work.
{}
\begin{center}
{\large\bf Figure Captions} \\[1cm]
\end{center}
\noindent
Fig.\ 1: For polytropic accretion with RH shocks, varitaion 
of normalized QPO frequencies (indicated by $\nu$ in the figure)
with specific flow energy ${\cal E}$ and angular momentum 
$\lambda$ for $\Phi_1$(A), $\Phi_2$(B), $\Phi_3$(C) and $\Phi_4$(D).
The figure is drawn for different $\gamma$ with the interval 
${\Delta}\gamma$=0.1.\\

\noindent
Fig.\  2: For polytropic accretion with RH shocks, 
varitaion 
of normalized QPO frequencies (indicated by $f_{norm}$ in the figure)
with shock location $r_{sh}$, post shock flow temperature $T_+$
and post shock integrated density $\Sigma_+$.\\

\noindent
Fig.\  3: For isothermal accretion with dissipative shocks,
varitaion 
of normalized QPO frequencies (indicated by $\nu$ in the figure)
with specific flow temperature in units of
$T_{10}=T/10^{10}$ Degree K (indicated by $\tau$ in the
figure) and angular momentum $\lambda$ for 
$\Phi_1$(A), $\Phi_2$(B), $\Phi_3$(C) and $\Phi_4$(D).
\\

\noindent
Fig.\  4: For isothermal accretion with dissipative shocks,
varitaion 
of normalized QPO frequencies (indicated by $f_{norm}$ in the figure)
with the shock location $r_{sh}$, shock compression 
ratio ${\Re}_{\Sigma}\left(=\frac{\Sigma_+}{\Sigma_-}\right)$ 
and the amount of energy dissipation at shock
${\Delta}{\epsilon}$.
\newpage
\plotone{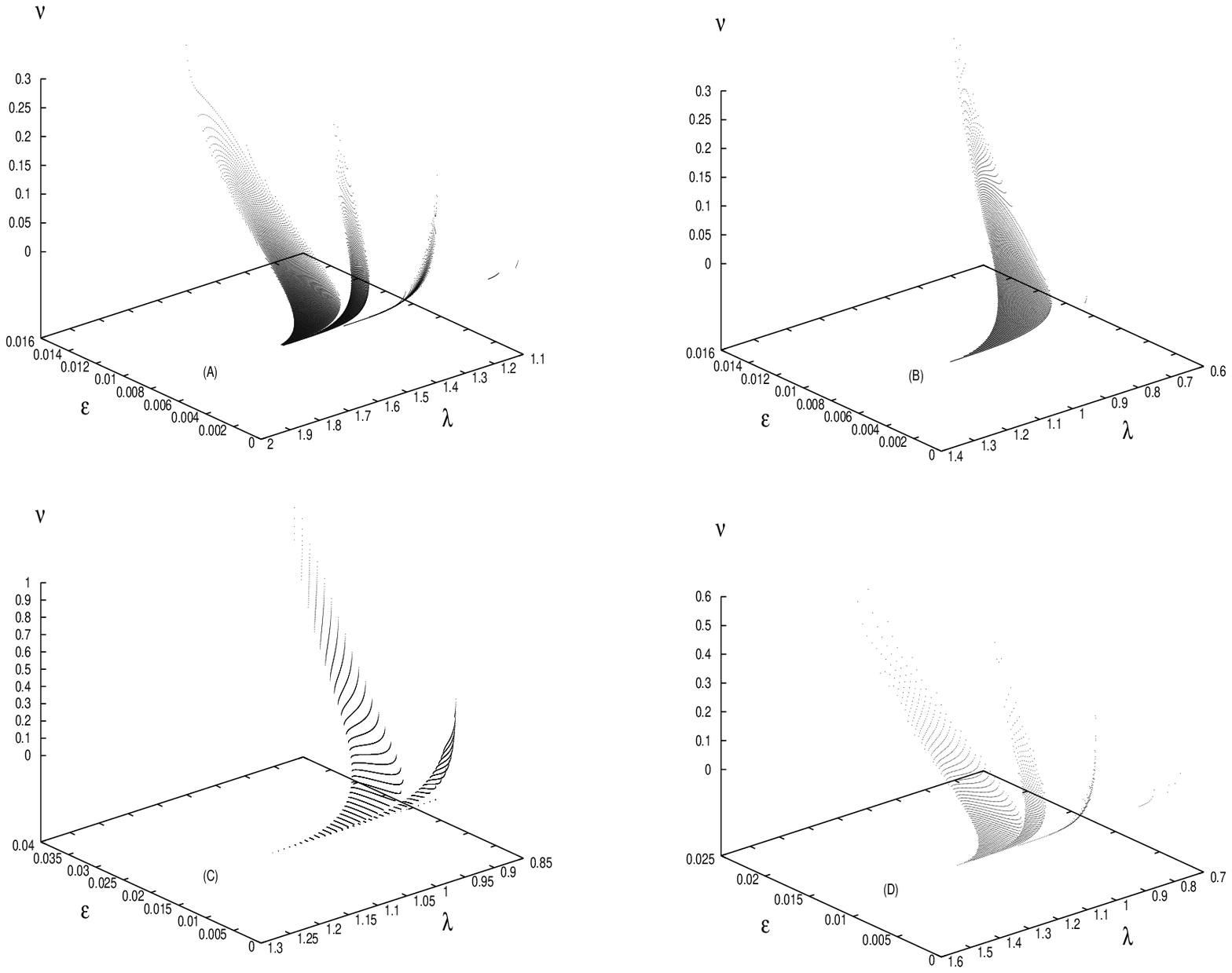}

\newpage
\plotone{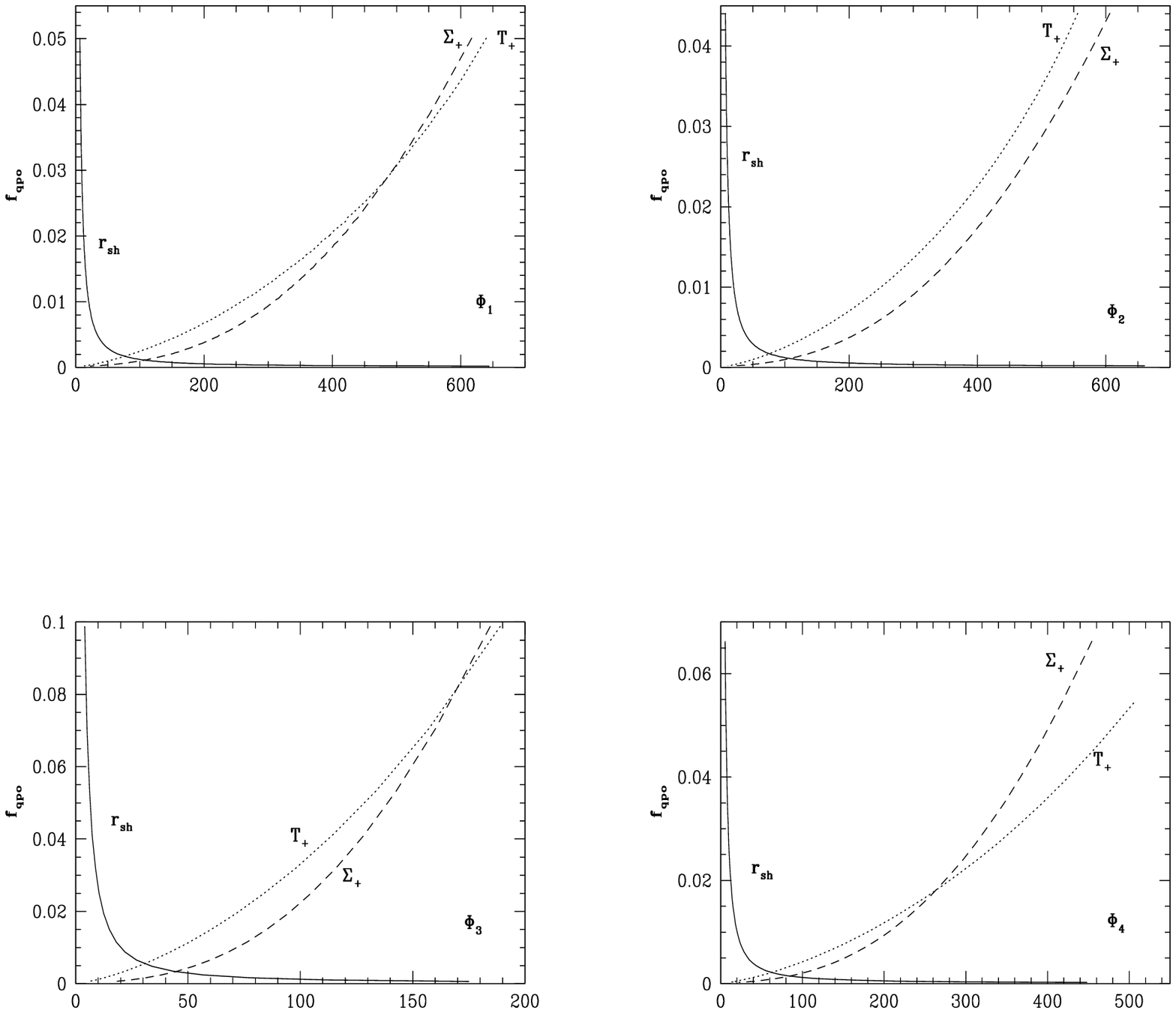}

\newpage
\plotone{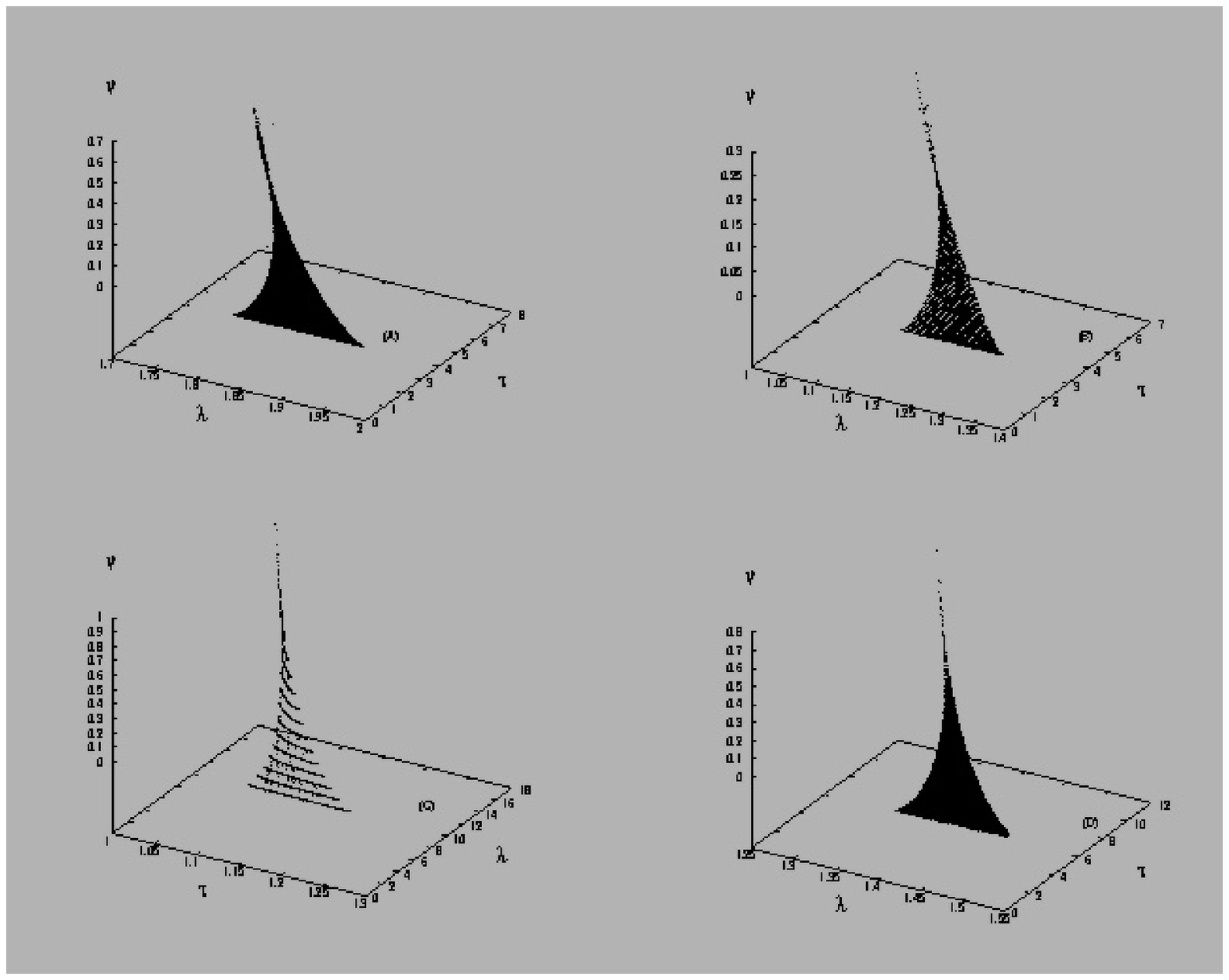}

\newpage
\plotone{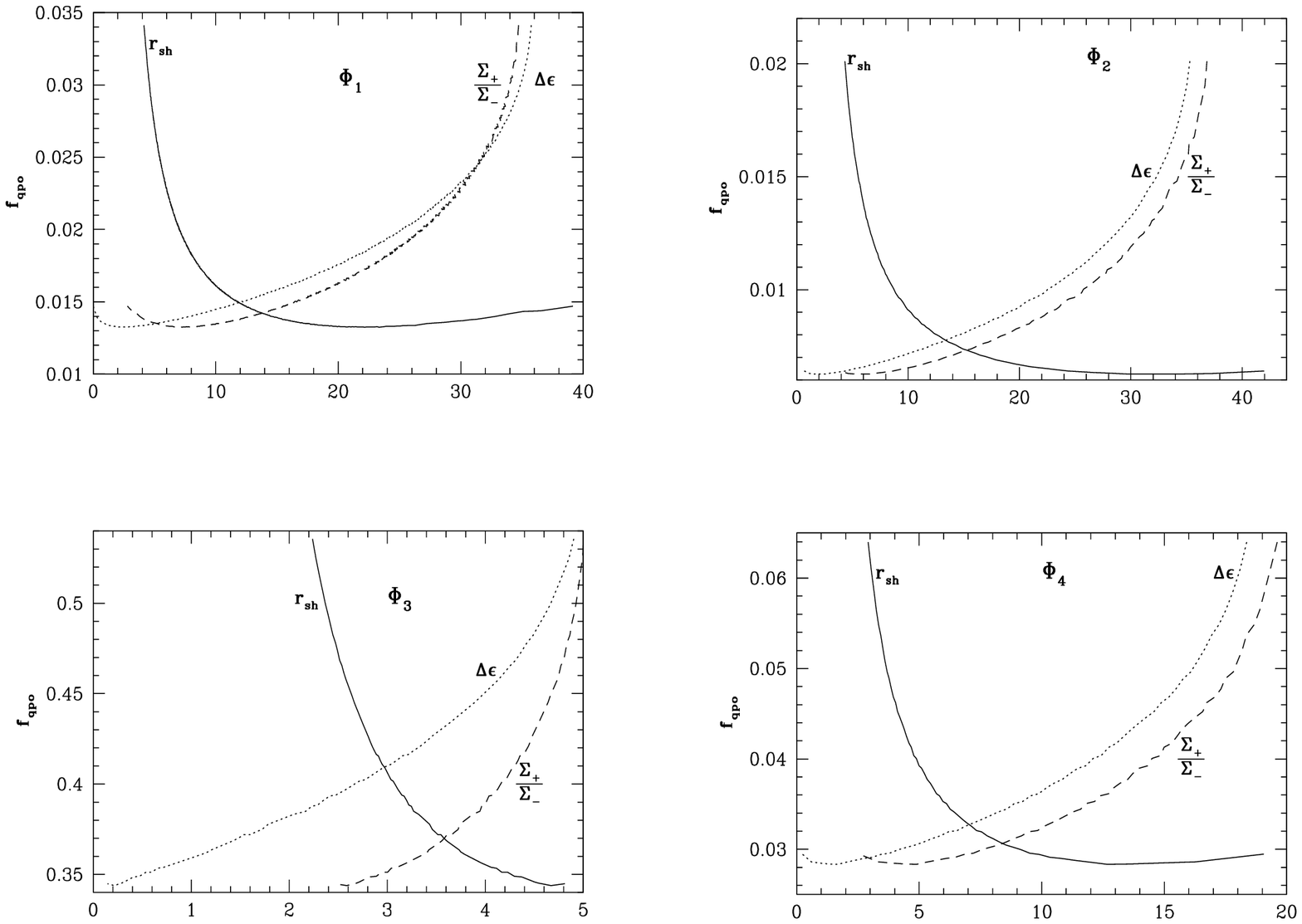}


\begin{thebibliography}{}
\bibitem[]{631} Artemova, I. V., Bj\"{o}rnsson,  G., \& Novikov, I. D.  1996, ApJ, 461, 565
\bibitem[]{633} Chakrabarti, S. K., \& Titarchuk, L. G. 1995 ApJ, 455, 623 (CT)
\bibitem[]{634} Chakrabarti, S. K., \& Manickam, S. G. 2000. ApJ, 531, L41 (CM)
\bibitem[]{635} Chen, X., \& Taam, R. E. 1995, Apj, 441, 354
\bibitem[]{637} Cui, W. et al. 1997, ApJ, 484, 383
\bibitem[]{638} Cui, W., 1999, in High Energy Processes in Accreting Black Holes, ASP Conference Series 161, ed. Juri Poutanen \& Roland Svensson. p.97
\bibitem[]{640} Das, T. K. 1998, in Observational Evidence for Black Holes in the Universe, Ed. S. K. Chakrabarti (Kluwer Academic: Holland),  p. 113
\bibitem[]{641} Das, T. K., 2002, ApJ, 577, 880 (D02)
\bibitem[]{642} Das, T. K., \& Chakrabarti, S. K. 1999, Class. Quantum Grav, 16, 3879
\bibitem[]{644} Das, T. K., Pendharkar, J., \& Mitra, S. 2003, ApJ, (DPM) (astro-ph/0301189)
\bibitem[]{645} Das, T. K., Rao, A. R., \& Vadawale, S.  2003, MNRAS, (astro-ph/0301344)
\bibitem[]{646} Fortner, B., Lamb, F. K., \& Miller, G. S. 
1989, Nature, 342, 775
\bibitem[]{648} Frank, J., King, A., \& Raine, D. 1992, Accretion Power in Astrophysics. 2nd. Edition. Cambridge University Press.
\bibitem[]{649} Kazanas, D., Hua, X.-M. \& Titarchuk, L. 1997, ApJ, 480, 735
\bibitem[]{} Matsumoto, R., Kato. S., Fukue. J., \& Okazaki. A. T.  1984, PASJ, 36, 71
\bibitem[]{650} Molteni, D., Sponholz, H., \& Chakrabarti, S. K. 1996, ApJ, 457, 805 (MSC)
\bibitem[]{651} Morgan, E. H., Remillard, R. A., \& Greiner,~J. 1997, ApJ, 482, 993
\bibitem[]{652} Muno, M. P., Morgan, E. H., \& Remillard, R. A. 1999, ApJ, 527, 321
\bibitem[]{653} Nowak, A. M., \& Wagoner, R. V. 1991, ApJ, 378, 656
\bibitem[]{654} Nowak, M. A., Wagoner, R. V., Begelman, M. C., \&
Lehr,D. E. 1997, ApJ, 477, L91
\bibitem[]{656} Paczy\'nski, B., \& Wiita, P. J. 1980, A \& A, 88, 23
\bibitem[]{657} Paul, B., Agrawal, P. C., Rao, A. R., Vahia, M. N., Yadav, J. S., Seetha, S.
\& Kasturirangan, K. 1998, ApJ, 472, L63
\bibitem[]{659} Remillard, R. et al. 1999, ApJ, 522, 397
\bibitem[]{660} Rutledge R E et al. 1999 ApJ Suppl. Ser. 124 265
\bibitem[]{662} Smith, D. M., Heindl, W. A., \& Swank, J. H. 2002, ApJ, 1, 362
\bibitem[]{663} Titarchuk, L., Lapidus, I., \& Muslimov, A. 1998, ApJ, 499, 315
\bibitem[]{664} van der Klis, M. 1995, in ``X--ray Binaries'', eds. W. H. G. Lewin, J. van Paradijs, \& E. P
. J. van den Heuvel (Cambridge U. Press, Cambridge) p. 252
\bibitem[]{666} Yadav, J. S., Rao, A. R., Agrawal, P. C., Paul, B., Seetha, S., \&
Kasturirangan, K. 1999, ApJ, 517, 935
\end{thebibliography}
\end{document}